\pdfoutput=1

\documentclass[11pt]{article}

\usepackage[final]{acl}

\usepackage{times}
\usepackage{latexsym}

\usepackage[T1]{fontenc}

\usepackage[utf8]{inputenc}

\usepackage{microtype}

\usepackage{inconsolata}
\usepackage{algorithm}
\usepackage{algorithmic}
\usepackage{subfigure}
\usepackage{makecell}

\usepackage{multirow}
\usepackage{makecell}
\usepackage{subfigure}
\usepackage{multicol}
\usepackage{multirow}
\usepackage{url}
\usepackage{amsmath,amsfonts,amsthm} 
\usepackage{graphicx}
\usepackage{subfigure}
 \usepackage{balance}
\usepackage{color}
\usepackage{tabularx}
\usepackage{array}
\usepackage{booktabs}
\usepackage{amssymb}
\usepackage{bbding}
\usepackage{bbm}
\usepackage{multirow}
\usepackage{makecell}
\usepackage{multicol}
\usepackage{multirow}
\usepackage{url}
\usepackage{amsmath,amsfonts,amsthm} 
\usepackage{balance}
\usepackage{color}
\usepackage{tabularx}
\usepackage{array}
\usepackage{booktabs}
\usepackage{amssymb}
\usepackage{bbding}
\usepackage[subfigure]{tocloft}
\title{Generative Cross-Modal Retrieval: Memorizing Images in Multimodal Language Models for Retrieval and Beyond}

\author{Yongqi Li$^{1}$, Wenjie Wang$^{2}$, Leigang Qu$^{2}$, Liqiang Nie$^{3}$, Wenjie Li$^{1}$,  Tat-Seng Chua$^{2}$ \\
        $^{1}$The Hong Kong Polytechnic University \\ 
        $^{2}$National University of Singapore
        $^{3}$Harbin Institute of Technology (Shenzhen)\\
        \texttt{\{liyongqi0,wenjiewang96,leigangqu,nieliqiang\}@gmail.com} \\ 
        \texttt{cswjli@comp.polyu.edu.hk dcscts@nus.edu.sg}}

\begin{document}
\maketitle
\begin{abstract}
The recent advancements in generative language models have demonstrated their ability to memorize knowledge from documents and recall knowledge to respond to user queries effectively. Building upon this capability, we propose to enable multimodal large language models (MLLMs) to memorize and recall images within their parameters. Given a user query for visual content, the MLLM is anticipated to ``recall'' the relevant image from its parameters as the response. Achieving this target presents notable challenges, including inbuilt visual memory and visual recall schemes within MLLMs. To address these challenges, we introduce a generative cross-modal retrieval framework, which assigns unique identifier strings to represent images and involves two training steps: learning to memorize and learning to retrieve. The first step focuses on training the MLLM to memorize the association between images and their respective identifiers. The latter step teaches the MLLM to generate the corresponding identifier of the target image, given the textual query input. By memorizing images in MLLMs, we introduce a new paradigm to cross-modal retrieval, distinct from previous discriminative approaches. The experiments demonstrate that the generative paradigm performs effectively and efficiently even with large-scale image candidate sets. 
\end{abstract}

\section{Introduction}
\label{sec:intro}
\begin{figure}[t]
\centering
\includegraphics[width=0.78\linewidth]{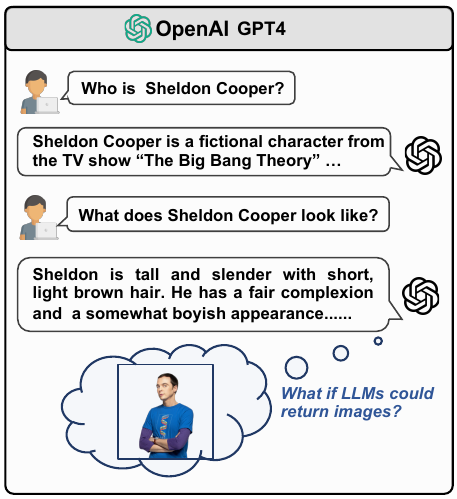}
\caption{Real cases from GPT4 illustrate the necessity of visual outputs for LLMs.}
\vspace{-1.5em}
\label{example}
\end{figure}

Recently, we have witnessed the explosive development of generative large language models (LLMs), such as GPT series~\cite{radford2019language,brown2020language} and LLaMA~\cite{touvron2023llama,touvron2023llama2}. Undergone extensive pretraining on document corpora and instruction tuning, these language models have demonstrated an impressive ability to memorize a lot of knowledge in their parameters and effectively recall them to answer users' instructions and queries. As shown in Figure~\ref{example}, GPT4\footnote{\url{https://openai.com/gpt-4}.} could directly respond to the user's question, ``Who is  Sheldon Cooper?'', without any external document or database. Building upon the advancements of LLMs, multimodal LLMs (MLLMs)~\cite{alayrac2022flamingo, li2023blip, liu2023visual,zhu2023minigpt,huang2023language} have been developed to expand the capabilities beyond text and allow users to express their needs using visual input. 

Despite the impressive capabilities of LLMs and MLLMs, their responses are limited to textual outputs. For instance, a user might ask, ``What does Sheldon Cooper look like?'' as shown in Figure~\ref{example}. While the MLLM tries to describe the person's appearance, it is often said that ``an image is worth a thousand words.'' It would greatly enhance the response capabilities of MLLMs if they could give visual outputs, like a photograph in this case.

A straightforward solution is to enhance MLLMs with external image synthesis tools, like diffusion models~\cite{dhariwal2021diffusion, ho2020denoising} and Generative Adversarial Networks~\cite{goodfellow2020generative}, for visual output capabilities. However, a significant challenge with these modules is their propensity to produce unrealistic or hallucinatory images, which cannot accurately describe real-world images, such as a photograph of ``Sheldon Cooper''. The integration of an image retrieval module~\cite{radford2021learning} seems a more viable solution. Nonetheless, such a combination often encounters a transition gap between two independent modules~\cite{lewis2020retrieval}. 
Considering the massive benefits of LLMs in memorizing textual knowledge, 
a bold and innovative idea emerges: 
Is it possible to equip MLLMs with the ability to memorize visual information within their parameters for retrieval and beyond? In this light, we formulate a generative cross-modal retrieval task: given a user query for visual content,
MLLMs are expected to recall desired images from their parameters directly as the response.

Accomplishing this task poses a significant challenge, necessitating the presence of two essential abilities of MLLMs: 1) Visual memory. As the prerequisite requirement, the MLLM model must possess the capability to memorize visual information within its parameters. This goes beyond simply encoding images into dense vectors within a vector database. It necessitates a distinct, differentiable, and integrated visual memory scheme within MLLMs' parameters. 2) Visual recall. Given a textual query, the MLLM should be able to recall the relevant visual information from the complicated visual memory bank. Above this, for user comprehension, the activated visual information must be grounded to the complete and original images rather than mere patches or fragmented visuals.

In this work, we propose a novel GeneRAtive Cross-modal rEtrieval framework, GRACE, to overcome the above issues. GRACE assigns images unique identifiers, where each identifier is a distinct string representing an image. Based on the identifiers, GRACE comprises two training steps, as illustrated in Figure~\ref{method}. 1) Learning to memorize. Given an image, the MLLM is trained to generate the corresponding identifier string via the standard text generation loss. The goal of this phase is for the MLLM to effectively learn and memorize the associations between the visual content of images and their respective identifiers. 2) Learning to retrieve. The MLLM is trained to generate the identifier string of the relevant image while given a textual query. In this way, the MLLM learns to associate user queries with visual memory. After the two training steps above, GRACE enables generative cross-modal retrieval: given a textual query, the MLLM generates an identifier string corresponding to a real image.

We delve into GRACE from various perspectives, including different identifier types, effectiveness, and efficiency of the generative paradigm. We evaluate GRACE on text-image matching datasets to verify the feasibility of generative cross-modal retrieval. Without any image's visual information during inference, GRACE performs comparably to the advance one-tower approaches (e.g., CLIP~\cite{radford2021learning}) and demonstrates higher efficiency with large-scale image sizes. It is acknowledged that as a new retrieval paradigm, GRACE still lags behind one-tower approaches. One-tower approaches are only applicable to ranking stage due to their low efficiency, while GRACE and CILP  are specifically designed for the retrieval stage. By comprehensive analysis, we hope to comprehensively understand its capabilities and limitations.

We believe exploring generative cross-modal retrieval holds great significance. 
\begin{itemize}
    \item Benefiting from inbuilt visual memory within MLLMs, GRACE introduces a new paradigm to cross-modal retrieval. GRACE transforms the original matching problem into a generation problem, eliminating the need for negative samples during training and retrieval index during inference. No matter the size of the image set, the retrieval efficiency remains constant. This new cross-modal retrieval paradigm leaves much room for investigation.
    \item Inbuilt visual memory serves for retrieval, yet its utility extends beyond mere retrieval. In Section~\ref{Beyond Cross-modal Retrieval}, we demonstrate that the MLLM could describe the memorized image and even answer questions about the memorized images, just like humans do. This opens up the possibility of injecting personalized visual experiences of humans into MLLMs for them to memorize and understand an individual's journey, and accomplish more visual tasks. 
\end{itemize}

\section{Related Work}
\subsection{Cross-modal Retrieval}
The current cross-modal retrieval (text-image matching) approaches can be categorized into the two frameworks and the one-tower framework based on how modality interaction is handled. One-tower framework~\cite{chen2020imram,diao2021similarity,lee2018stacked,qu2021dynamic} embraces fine-grained cross-modal interactions to achieve matching between fragments (e.g., objects and words). 
As for the two-tower framework~\cite{chen2021learning,faghri2017vse++,zheng2020dual,qu2020context}, images and texts are independently mapped
into a joint feature space in which the semantic similarities are calculated via cosine function or Euclidean distance. Both the one-tower framework and the two-tower framework formulate the cross-modal retrieval as a discriminative problem, which relies on discriminative loss and negative samples to learn an embedding space. In this work, we explore a new generative paradigm for cross-modal retrieval.

\subsection{Generative Retrieval}
Generative retrieval is an emerging new retrieval paradigm in text retrieval, which generates identifier strings of passages as the retrieval target. Instead of generating entire passages, this approach uses identifiers to reduce the amount of useless information and make it easier for the model to memorize and learn~\cite{li2023multiview}. Different types of identifiers have been explored in various search scenarios, including passage titles (Web URLs), numeric IDs, and substrings of passages, as shown in previous studies~\cite{de2020autoregressive, tay2022transformer,bevilacqua2022autoregressive,li2023learning,zhang2023irgen,li2023generative}. Generative retrieval gains a lot of attention in text retrieval, as it could take advantage of the powerful generative language models. However, how to facilitate cross-modal retrieval in a generative way is still an untapped problem.

\begin{figure*}[t]
\centering
  \includegraphics[width=1.0\linewidth]{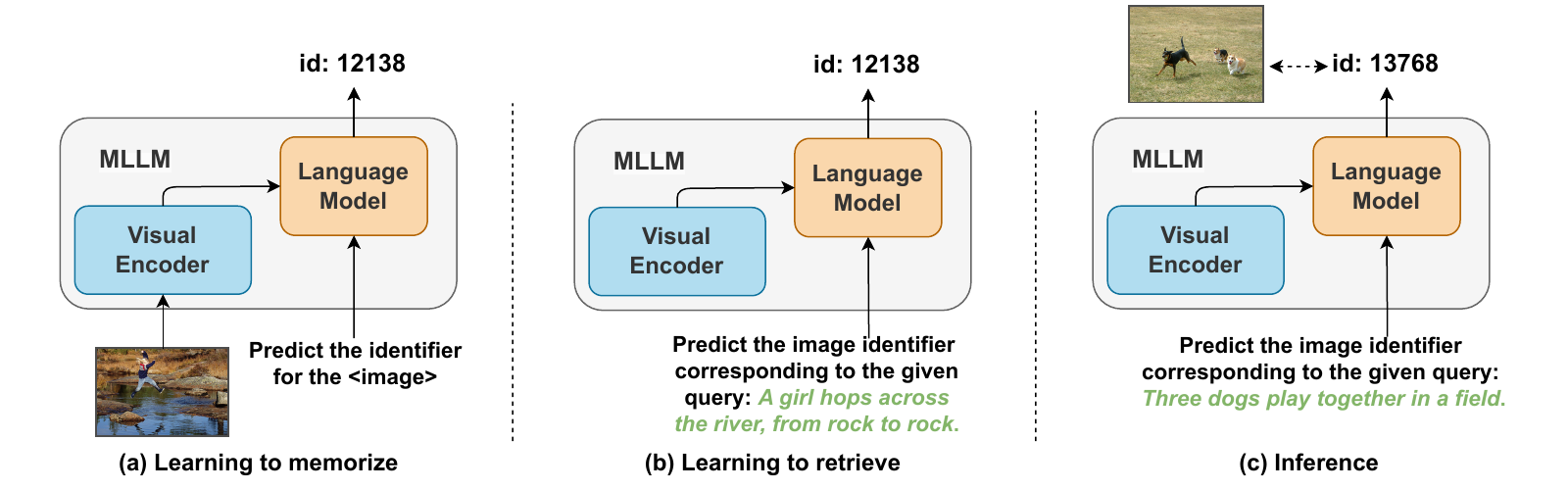}
  \vspace{-1.5em}
  \caption{Illustration of our proposed generative cross-modal framework, GRACE, which involves two training steps. (a) Learning to memorize: GRACE trains an MLLM model to memorize images into its parameters. (b) Learning to retrieve: GRACE trains the model to generate the target image's identifiers given queries. (c) Inference: The MLLM directly generates identifiers as the retrieval results.}
  \vspace{-1em}
  \label{method}
\end{figure*}

\subsection{Multimodal  Language Model}
We have witnessed the explosive development of generative language models, such as GPT~\cite{radford2019language,brown2020language} and LLaMA~\cite{touvron2023llama}, that demonstrate remarkable capabilities in instruction following and in-context learning. Building upon the advancements of LLMs, MLLMs~\cite{alayrac2022flamingo, li2023blip, liu2023visual,zhu2023minigpt,huang2023language} have been developed to enable LLMs to process images as input. Despite the success of MLLMs in various vision-language tasks, they currently lack the ability to unify cross-modal retrieval into their application. In this work, we propose a generative cross-modal retrieval framework that empowers MLLMs to retrieve relevant images from their parameters given textual queries.
\section{Method}
\subsection{Preliminary}
\label{Preliminary}
\textbf{Task definition}. Generative cross-modal retrieval defines new requirements, i.e., removing visual input during inference, for cross-modal retrieval, but could be evaluated with original cross-modal tasks. Text-to-image retrieval aims to retrieve relevant images from a database $\mathcal{D}_{I}$ when given a textual query $q$.  

\textbf{Multimodal language model}. As our method is conducted based on multimodal language models, it is essential to give relevant background of multimodal language models. Multimodal language models could be regarded as generative language models that incorporate image inputs, including GPT4V\footnote{\url{https://openai.com/research/gpt-4v-system-card}.}, BILP~\cite{li2023blip}, flamingo~\cite{alayrac2022flamingo}, and Kosmos~\cite{huang2023language}. Considering factors including convenience and model sizes, we have chosen Flamingo as the backbone for our method and took the open-flamingo implementation~\cite{awadalla2023openflamingo}.

Flamingo consists of three main components: a generative language model, a visual encoder, and cross-attention layers. The visual encoder is responsible for extracting patch features from the input images. The generative language model receives text input that includes a special token, ``\textless image\textgreater'', which indicates the presence of an image. Through the cross-attention layers, the ``\textless image\textgreater'' token could attend to the patch features extracted by the visual encoder. This allows Flamingo to predict the next text token based on all previous text tokens and the most recent image. For more detailed information, please refer to the original paper on Flamingo.

\subsection{Overview}
\label{Overview}
In this work, we present GRACE, a novel generative cross-modal retrieval framework, as illustrated in Figure~\ref{method}. As previously discussed, addressing the challenges of visual memory and visual recall is essential for generative cross-modal retrieval. Towards this objective, GRACE assigns \textbf{unique} identifiers to images in the dataset $\mathcal{D}_{I}$. This strategy allows the model to learn mappings from images to their respective identifiers, facilitating visual memory. Moreover, the model could generate identifiers as retrieval results rather than generate real images. Representing images as identifiers underpins our training scheme, which is divided into two core steps: ``learning to memorize'' and ``learning to retrieve''. The two training steps are designed to enable the model to effectively memorize images in parameters and subsequently learn to recall them in response to textual queries.

\begin{figure}[t]
\centering
\includegraphics[width=1.0\linewidth]{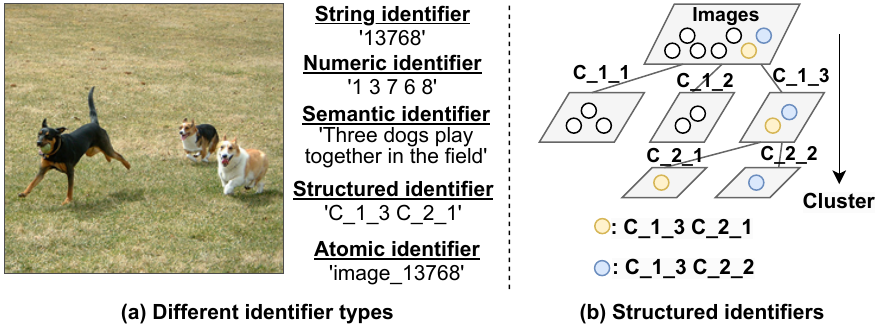}
\vspace{-2em}
\caption{(a) depicts an image accompanied by various identifier types. (b) shows the formation of structured identifiers, where each image's identifier is represented as its unique path within a cluster tree.}
\vspace{-1em}
\label{identifiers}
\end{figure}

\subsection{Image Identifiers}
\label{sec: Image Identifiers}
Image identifiers are crucial for the whole framework, and we explore the following different types of identifiers:

 \textbf{String identifier}. We randomly shuffle the images in $\mathcal{D}_{I}$, and assign them digital numbers ranging from $1$ to $|\mathcal{D}_{I}|$. It is noted that the digital numbers are represented as strings in MLLMs and may be tokenized into multiple tokens determined by the tokenizer. For instance, an image may be assigned the identifier ``13768'' and tokenized into two tokens: ``13'' and ``768''.
 
 \textbf{Numeric identifier}.  Similar to the string identifier, the numeric identifier ranges from $1$ to $|\mathcal{D}_{I}|$. However, we include spaces in the numeric identifier, resulting in the tokenization into individual digits. For example, an image with the identifier ``1 3 7 6 8'' will be tokenized into the sequence of tokens ``1'', ``3'', ``7'', ``6'', and ``8''. It is worth noting that the numeric identifier only utilizes ten tokens from the vocabulary to represent images, but the sequence length is typically longer than that of the string identifier.
 
\textbf{Semantic identifier}. Since the identifiers are utilized to represent images,  image captions that describe the content of images can be considered as identifiers. These image captions are naturally token sequences that can be learned by multimodal language models. Some images in $|\mathcal{D}_{I}|$ belong to the test set, and their captions should not be utilized. To avoid data leaks, we train an image caption model based on the training set and generate captions for the images in the test set as their identifiers. 

\textbf{Structured identifier}. We assign structure identifiers to images using an unsupervised clustering approach. We utilize the image encoder in CLIP to obtain the embeddings of images. Subsequently, we apply the k-means algorithm~\cite{ahmed2020k} to cluster these embeddings, resulting in all images being grouped into $k$ clusters. Each document is then assigned an identifier based on the number of their cluster IDs. For clusters that contain more than a certain number of documents (denoted as $c$), we recursively apply the algorithm~\cite{tay2022transformer}. In this process, the identifier of the next level is appended to the existing identifier, forming a hierarchical structure. We represent each cluster using special tokens, such as "C\_1\_3", which indicates the third cluster in the first level. These special tokens are added to the token vocabulary of the multimodal language model. Similar images tend to have similar structured identifiers, meaning they have similar paths in the cluster tree.

\textbf{Atomic identifier}. We assign a dedicated token as its identifier to identify each image uniquely. We expand the token vocabulary by introducing new tokens to ensure compatibility with the existing tokens. Each image is then assigned a special token, such as "I\_13768", which is a complete token in the vocabulary and will not be further tokenized into sub-tokens. This approach allows us to avoid any conflicts with the original tokens while providing a distinct identifier for each image.

We present the various types of identifiers for the same image in Figure~\ref{identifiers}, highlighting their distinct characteristics. It is evident that different identifier types possess different attributes. String, numeric, and atomic identifiers do not provide any prior knowledge about the image content, whereas semantic and structured identifiers do. Furthermore, the use of structured and atomic identifiers necessitates the inclusion of new tokens in the vocabulary, whereas the other identifier types do not require such modifications.
\subsection{Learning to Memorize}
\label{sec: Learning to Memorize}
We have represented images in the dataset $\mathcal{D}_I$ using unique identifiers, that is, as a sequence of tokens. Then we train a multimodal language model, denoted as \textbf{MLLM}, to encapsulate these images within its parameters. Specifically, for an image $i \in \mathcal{D}_I$, we train the model to associate this image with its corresponding identifier, denoted as $\mathcal{I}$. This process is formulated as follows:
\begin{equation}  \label{eq:identifier_mapping}
   \mathcal{I} = \textbf{MLLM}(i; \text{inst-m}),
\end{equation}
where $\text{inst-m}$ is a textual instruction given as ``Predict the identifier for the \texttt{<image>}''. Here, ``\texttt{<image>}'' is a placeholder token in Flamingo, designed to focus on the visual features of the input. This learning to memorize step allows the model to learn the mappings from visual inputs to their corresponding identifiers, to effectively encode image-level visual memories within its parameters.
\subsection{Learning to Retrieve}
\label{sec: Learning to Retrieve}
Merely memorizing images within its parameters is insufficient for the MLLM. The model must be capable of recalling the corresponding images in response to users' queries. To achieve this, we train the MLLM to predict the appropriate identifier when given a specific query $q$. This process is outlined as follows:
\begin{equation}  \label{eq:query_identifier_mapping}
   \mathcal{I} = \textbf{MLLM}(q; \text{inst-r}),
\end{equation}
where $\text{inst-r}$ is a textual instruction, ``Predict the image identifier corresponding to the given query''.
\subsection{Inference}
\label{sec: Inference}
Post-training, the MLLM model could retrieve images akin to text generation. The process involves inputting a query into the MLLM, and then the model predicts several identifier strings through beam search. Since each identifier uniquely corresponds to an image, the generation results are the retrieval results.
\begin{table*}[t]
\renewcommand\arraystretch{1}
  \centering
    \scalebox{1.0}{
    \begin{tabular}{ccccccccc}
    \toprule
  \multirow{2}*{ Paradigm}& \multicolumn{1}{c}{\multirow{2}*{Methods}}
    &\multicolumn{3}{c}{\makecell[c]{Flickr30K}}&&\multicolumn{3}{c}{\makecell[c]{MS-COCO (5K)}}\\\cline{3-5}\cline{7-9}
         &&R@1&R@5&R@10&&R@1&R@5&R@10\cr
    \toprule
    \multicolumn{1}{c}{\multirow{4}*{\makecell[c]{Two-tower}}}&VSE++ (\citeauthor{faghri2017vse++}) &39.6 &70.1 &79.5&&30.3 &59.4 &72.4 \cr
    &Dual-path (\citeauthor{zheng2020dual}) &39.1 &69.2 &80.9&&25.3 &53.4 &66.4 \cr
    &CAMERA (\citeauthor{qu2020context})&\underline{58.9} &\underline{84.7} &\underline{90.2}&&\underline{39.0} &\textbf{70.5} &\textbf{81.5} \cr
    &CLIP (\citeauthor{radford2021learning})&58.4&81.5&88.1&&37.8&62.4&72.7 \cr \toprule
    \multicolumn{1}{c}{\multirow{5}*{GRACE}}&Numeric Identifier&22.5&28.9&29.4&&0.03&0.14&0.28\cr 
    &String Identifier&30.5&39.0&40.4&&0.12&0.37&0.88\cr 
    &Semantic Identifier&22.9&34.9&37.4&&13.3&30.4&35.9\cr 
    &Structured Identifier&37.4&59.5&66.2&&16.7&39.2&50.3\cr
    &Atomic Identifier&\textbf{68.4}&\textbf{88.9}&\textbf{93.7}&&\textbf{41.5}&\underline{69.1}&\underline{79.1}\cr
\toprule
    \end{tabular}}
    \vspace{-0.5em}
    \caption{Performance of text-to-image retrieval on Flickr30K and MS-COCO (5K) datasets. The best results in each group are marked in Bold, while the second-best ones are underlined. One-tower approaches demonstrate superior performance on the two datasets, but they are not considered as baselines due to their high computational overhead, which makes them impractical for the retrieval stage.}  \label{tab:Retrieval performance}
    \vspace{-1em}
\end{table*}

\textbf{Constrained generation}. To confine the generation to within-corpus results and ensure they fall within the test set, we implement constrained beam search in the MLLM. This approach leverages a Trie, a form of k-ary search tree, for efficient key location within a set. Specifically, we store all image identifiers into the Trie. The Trie structure, upon receiving a prefix string, suggests potential tokens found in the identifiers. This mechanism ensures that every generated identifier accurately matches an existing image's identifier. Furthermore, we employ beam search~\cite{sutskever2014sequence}, a widely-used technique, for generating multiple identifiers concurrently. These identifiers are each assigned a language model score, facilitating the creation of a ranked list based on these scores. Consequently, the ranked identifiers correspond to a ranked list of images.
\section{Experiments}
\subsection{Datasets and Baselines}
We evaluated our proposed generative cross-modal retrieval framework, GRACE, on two commonly-used datasets: Flickr30K~\cite{young2014image} and MS-COCO~\cite{lin2014microsoft}. Flickr30K contains 31,783 images sourced from Flickr. Each image is associated with five human-annotated sentences. We adopted the data split used by ~\citeauthor{li2019visual}, comprising 29,783 images for training, 1,000 for validation, and 1,000 for testing. MS-COCO comprises 123,287 images, and each MS-COCO image comes with five sentences of annotations. We followed the dataset split proposed in ~\cite{lee2018stacked}, utilizing 113,287 images for training, 5,000 for validation, and 5,000 for testing. Consistent with prior studies~\cite{young2014image, chen2021learning}, we evaluated our method using the standard recall metric $R@K$ where $K$ is set to 1, 5, and 10.

Considering the efficiency and 
applicability, we compared GRACE with two-tower approaches, including VSE++~\cite{faghri2017vse++}, Dual-path~\cite{zheng2020dual}, CAMERA~\cite{qu2020context}, and CLIP~\cite{radford2021learning}, as our baseline models. \textbf{One-tower approaches usually have heavy computational overhead, focusing on the ranking stage rather than the retrieval stage. Therefore, we did not include them as baselines.}

\textbf{Implement Details} are detailed in Appendix~\ref{sec: Implement Details}. 

\subsection{Overall Results}
The summarized comparisons are presented in Table~\ref{tab:Retrieval performance}. Analysis of this table led to the following observations: 1) GRACE demonstrated the capability to recall relevant images in response to textual queries without input of image content. This underscores the feasibility of generative cross-modal retrieval. 2) We also noticed variability in performance among GRACE with different identifiers. Specifically, numeric and string identifiers yielded very low performance on the MS-COCO dataset. This poor performance can be attributed to the lack of pre-knowledge provided by these identifiers to the MLLM. The inconsistent correlation between similar images and their identifiers makes it challenging for the MLLM to memorize and establish accurate relationships, especially as the dataset size increases. Furthermore, numeric identifiers underperform string identifiers, likely due to their requirement for more generation steps, which increases the chance of errors. 3) In contrast, semantic identifiers, which are based on the image's content, showed better results than numeric and string identifiers. However, their effectiveness was somewhat limited due to the minimal differentiation among semantic identifiers for different images. This was particularly problematic in cases where images shared the same captions, causing the model to generate semantically correct but contextually incorrect identifiers. 4) Structured identifiers achieved good performance by effectively utilizing the image's embedding information through a clustering approach. This hierarchical structure significantly enhanced the MLLM's ability to memorize all images in the dataset. 5) Finally, atomic identifiers were found to be the most effective, even outperforming the CLIP model. This approach assigns a unique token in the vocabulary for each image, ensuring distinct identification. However, this method also has its challenges, as increasing the number of images directly enlarges the vocabulary size of the MLLM, potentially impacting scalability.

These findings highlight the importance of identifier types in generative cross-modal retrieval and shed light on the trade-offs involved in different approaches.

\begin{table}[t]
\renewcommand\arraystretch{1}
  \centering
  \setlength{\tabcolsep}{4mm}{
\resizebox{\linewidth}{!}{
    \begin{tabular}{cccc}
    \toprule
    \multicolumn{1}{c}{\multirow{2}*{GRACE}}
    &\multicolumn{3}{c}{\makecell[c]{Flickr30K}}\\\cline{2-4}
         &R@1&R@5&R@10\cr
    \toprule
    Numeric Identifier&22.5&28.9&29.4\cr
    w/o learning to memorize&18.2&24.3&24.9\cr
    w/o constrained generation&7.72&16.7&21.1\cr\toprule
    String Identifier&30.5&39.0&40.4\cr
    w/o learning to memorize&26.1&33.3&34.6\cr
    w/o constrained generation&10.9&22.3&28.0\cr\toprule
     Semantic Identifier&22.9&34.9&37.4\cr
    w/o learning to memorize&19.3&31.2&34.3\cr
    w/o constrained generation&0.6&2.3&3.0\cr\toprule
    Structured Identifier&37.4&59.5&66.2\cr
    w/o learning to memorize&36.5&61.1&68.2\cr
    w/o constrained generation&10.2&22.3&29.3\cr\toprule    
    \end{tabular}}}  
    \vspace{-0.5em}
    \caption{ Ablation study results for GRACE. The term ``w/o learning to memorize'' indicates the omission of the ``learning to memorize'' training step, and ``w/o constrained generation'' refers to free generation without any restriction during the inference stage.}
    \vspace{-1em}
    \label{tab:ablation study}
\end{table}
\subsection{Ablation Study}
Our approach integrates two key training steps: learning to memorize and learning to retrieve. Does the ``learning to memorize'' phase significantly enhance retrieval performance? During the inference stage, we employed constrained generation to ensure the prediction of valid identifiers. How crucial is constrained generation to the overall retrieval process? To address these questions, we performed experiments by selectively omitting the ``learning to memorize'' step and the constrained generation process. The outcomes of these experiments are detailed in Table~\ref{tab:ablation study}.

In our experiments, we observed a slight decrease in performance when the ``learning to memorize'' training step was removed. This suggests that while important, this step is not the sole contributor to effective retrieval. Intriguingly, the ``learning to retrieve'' phase can be considered another form of memorization, where the model focuses on the image's description rather than its visual content. As a result, the model retains some capability to recall correct images even without the ``learning to memorize'' step. However, a significant decline in performance was noted upon removing the constrained generation step. This can be attributed to two primary factors. (1) Generation of out-of-corpus identifiers: without constrained generation, the model tends to predict identifiers that do not correspond to any image in the corpus. This issue is especially pronounced with semantic identifiers, where the model may generate any textual description, leading to inaccurate retrieval. (2) Prediction of identifiers belonging to the training set. For other types of identifiers, while the model still predicts special tokens corresponding to these identifiers, it often predicts images in the training set. The vast number of images in the training set could also be relevant to the given textual query, significantly increasing the difficulty of recalling the correct image in the test set.
\begin{figure}[t]
\centering
\includegraphics[width=0.8\linewidth]{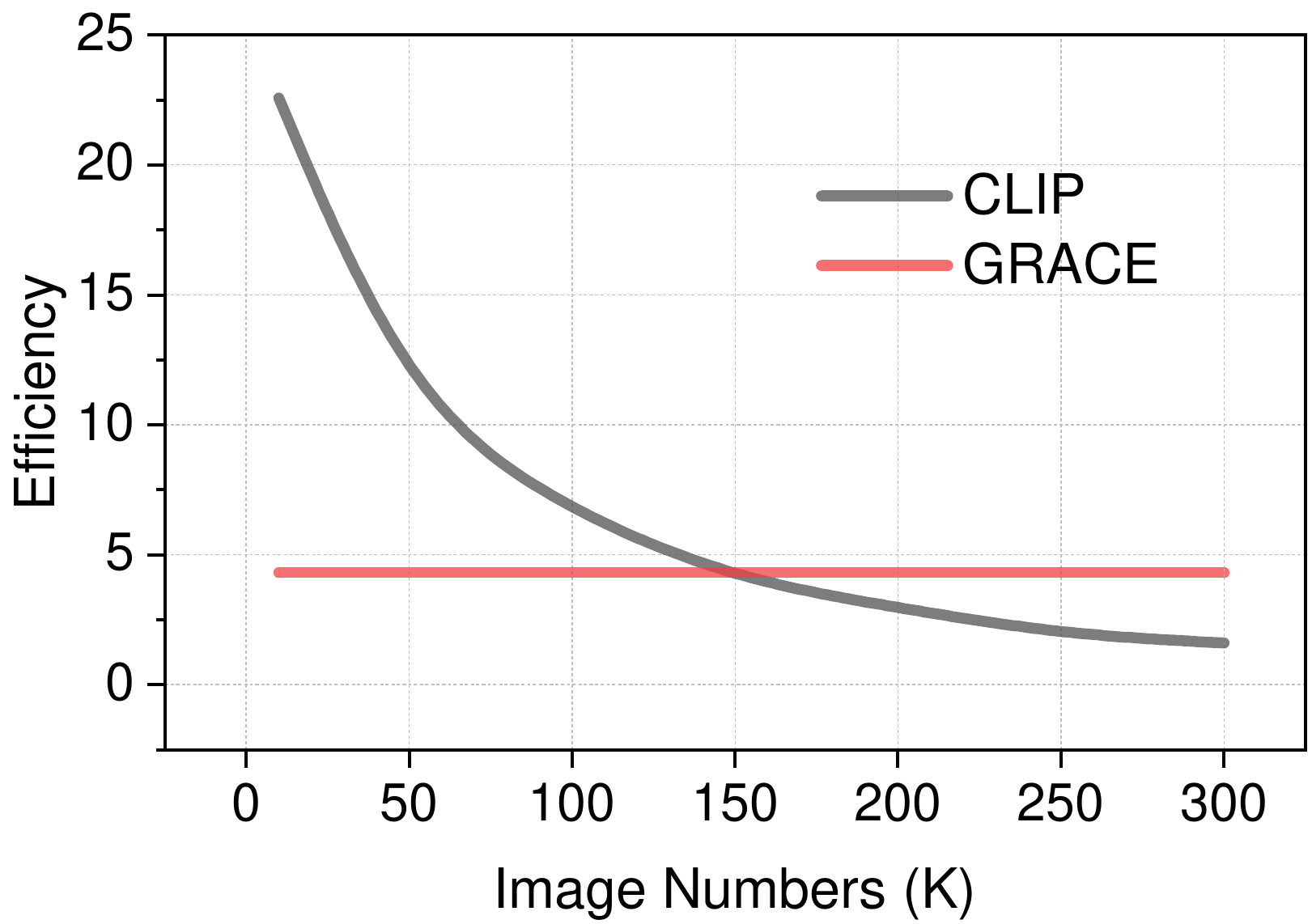}
\vspace{-0.5em}
\caption{The efficiency of {CLIP} and {GRACE} varies with image size, measured in terms of queries processed per second. As the image size increases, {GRACE} demonstrates superior efficiency.}
\vspace{-1em}
\label{efficiency}
\end{figure}

\begin{figure*}[t!]
\centering
\includegraphics[width=1.0\linewidth]{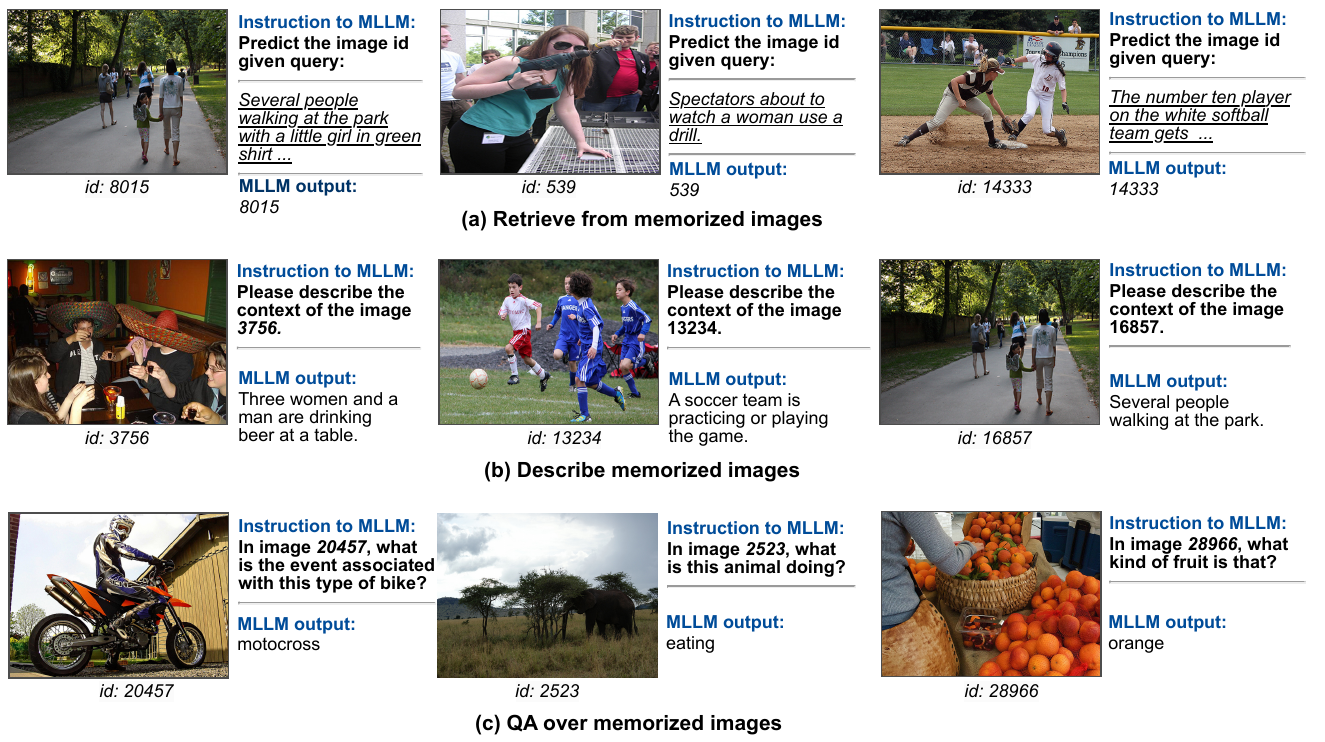}
\vspace{-1em}
\caption{Cases of interaction with memorized images for an MLLM include retrieving the memorized images, describing them, and answering questions about them, based on specific instructions. It is noted that the MLLM model responds to user instructions without any image input, relying solely on memorized visual information.}
\vspace{-1em}
\label{cases}
\end{figure*}

\subsection{Efficiency Analysis}

In large-scale cross-modal retrieval, efficiency emerges as a crucial factor. This is why the one-tower framework, effective for small-scale ranking, falls short in the retrieval stage. To address this, we conducted experiments comparing the efficiency of CLIP and GRACE. CLIP can pre-encode all images into vectors, incurring most of its inference cost from text encoding and calculating the similarity between text embeddings and image embeddings. In contrast, the generative framework necessitates generating identifiers. We assessed the query latency of both CLIP and GRACE to varying image sizes, with detailed results presented in Figure~\ref{efficiency}.

Our findings are insightful. Firstly, CLIP’s inference speed decreases progressively as image size increases, owing to the escalating number of similarity calculations required. Secondly, the inference speed of our generative framework remains nearly constant, a result of encoding all images directly into its parameters. Thirdly, when image sizes exceed a certain threshold (about 150,000 images), our generative framework surpasses CLIP in terms of inference speed, and this advantage grows as image sizes continue to increase. Lastly, these findings underscore that the generative framework is not only capable of large-scale image retrieval but can also perform comparably to two-tower approaches.

\subsection{Beyond Cross-modal Retrieval}
\label{Beyond Cross-modal Retrieval}
We enable the MLLM to memorize images within its parameters using unique identifiers. Once the images are adequately memorized, the MLLM can produce the corresponding images (identifiers) to respond to users' queries, as illustrated in Figure~\ref{cases} (a).

While the visual memory in the MLLM facilitates image retrieval, its applications are not restricted to retrieval alone after other instruction tunings. We present two examples in Figure~\ref{cases} (b) and Figure~\ref{cases} (c), respectively.
\begin{itemize}
\item \textbf{Describing memorized images}: As the MLLM has successfully memorized certain images, it is capable of providing a description of the image's content when prompted. As depicted in the examples shown in Figure~\ref{cases}, when given an instruction such as ``please describe the context of the image \textit{3,756}'', the model is able to provide a description of the image, albeit not in great detail.
\item \textbf{QA over memorized images}: Similarly, the model is capable of answering some questions over the memorized images. Given an instruction consisting of the image identifier and question, the model can answer based solely on memorization without any image input.
\end{itemize}
\subsection{Beam Size Analysis}
We conducted experiments to analyze the beam size of GRACE, as detailed in Appendix~\ref{sec: Beam size analysis}.
\section{Conclusion and Future Work}
In this paper, we delved into a novel memorization mechanism for the MLLM to memorize images within its parameters. Building upon inbuilt visual memory within MLLM, we proposed a generative cross-modal retrieval framework, which introduces a fresh paradigm in cross-modal retrieval. This paradigm transforms the original matching problem into a generation problem, eliminating the need for negative samples during training and image indexing during inference. Our experiments demonstrate that the generative paradigm performs effectively and efficiently even with large-scale image sizes. Furthermore, we showcased the MLLM's ability to interact (e.g., describe and QA) with memorized images, following specific instructions.

Moving forward, we aim to further develop this topic from the following perspectives. On the one hand, although our generative framework achieves comparable performance to previous cross-modal retrieval approaches, there are still challenges to address, such as the limitations of current identifiers. Exploring more effective identifiers, like ``visual tokens~\cite{van2017neural}'', would help to enhance generative cross-modal retrieval further. 
On the other hand, since we have enabled MLLMs to memorize and interact with images, it opens up the possibility of injecting personalized visual experiences of humans into MLLMs for them to understand an individual's visual journey and accomplish more visual tasks. 
\section*{Limitations}
This work introduces a new paradigm in text-image retrieval, but it also has some limitations to be addressed. 1) The evaluation of GRACE's image retrieval ability on Flickr30K and MS-COCO was compared with two-tower baselines. However, it is important to note that Flickr30K and MS-COCO are also used as benchmarks for text-image ranking approaches, where one-tower frameworks have dominated. This may confuse newcomers to the field, as they may perceive GRACE and two-tower approaches as lagging behind the one-tower framework. However, it should be noted that GRACE and one-tower approaches focus on image retrieval, placing high demands on retrieval efficiency, while two-tower approaches are primarily suitable for the ranking stage, allowing for more time-consuming calculations to improve performance. 2) The identifiers currently used by GRACE are not as satisfactory as expected, only yielding results comparable to previous methods. However, as a pioneering work, the main significance of this work lies in validating the feasibility of generative cross-model retrieval. Further research is expected to enhance this paradigm.

\section*{Ethics Statement}
The datasets used in our experiment are publicly released and labeled through interaction with humans in English. In this process, user privacy is protected, and no personal information is contained in the dataset. The scientific artifacts that we used are available for research with permissive licenses. And the use of these artifacts in this paper is consistent with their intended use. Therefore, we believe that our research work meets the ethics of ACL. 
\bibliography{acl_latex}

\begin{thebibliography}{37}
\expandafter\ifx\csname natexlab\endcsname\relax\def\natexlab#1{#1}\fi

\bibitem[{Ahmed et~al.(2020)Ahmed, Seraj, and Islam}]{ahmed2020k}
Mohiuddin Ahmed, Raihan Seraj, and Syed Mohammed~Shamsul Islam. 2020.
\newblock The k-means algorithm: A comprehensive survey and performance evaluation.
\newblock \emph{Electronics}, 9(8):1295.

\bibitem[{Alayrac et~al.(2022)Alayrac, Donahue, Luc, Miech, Barr, Hasson, Lenc, Mensch, Millican, Reynolds et~al.}]{alayrac2022flamingo}
Jean-Baptiste Alayrac, Jeff Donahue, Pauline Luc, Antoine Miech, Iain Barr, Yana Hasson, Karel Lenc, Arthur Mensch, Katherine Millican, Malcolm Reynolds, et~al. 2022.
\newblock Flamingo: a visual language model for few-shot learning.
\newblock \emph{Advances in Neural Information Processing Systems}, 35:23716--23736.

\bibitem[{Awadalla et~al.(2023)Awadalla, Gao, Gardner, Hessel, Hanafy, Zhu, Marathe, Bitton, Gadre, Sagawa et~al.}]{awadalla2023openflamingo}
Anas Awadalla, Irena Gao, Josh Gardner, Jack Hessel, Yusuf Hanafy, Wanrong Zhu, Kalyani Marathe, Yonatan Bitton, Samir Gadre, Shiori Sagawa, et~al. 2023.
\newblock Openflamingo: An open-source framework for training large autoregressive vision-language models.
\newblock \emph{arXiv preprint arXiv:2308.01390}.

\bibitem[{Bevilacqua et~al.(2022)Bevilacqua, Ottaviano, Lewis, Yih, Riedel, and Petroni}]{bevilacqua2022autoregressive}
Michele Bevilacqua, Giuseppe Ottaviano, Patrick Lewis, Wen-tau Yih, Sebastian Riedel, and Fabio Petroni. 2022.
\newblock Autoregressive search engines: Generating substrings as document identifiers.
\newblock \emph{arXiv preprint arXiv:2204.10628}.

\bibitem[{Brown et~al.(2020)Brown, Mann, Ryder, Subbiah, Kaplan, Dhariwal, Neelakantan, Shyam, Sastry, Askell et~al.}]{brown2020language}
Tom Brown, Benjamin Mann, Nick Ryder, Melanie Subbiah, Jared~D Kaplan, Prafulla Dhariwal, Arvind Neelakantan, Pranav Shyam, Girish Sastry, Amanda Askell, et~al. 2020.
\newblock Language models are few-shot learners.
\newblock \emph{Advances in neural information processing systems}, 33:1877--1901.

\bibitem[{Chen et~al.(2020)Chen, Ding, Liu, Lin, Liu, and Han}]{chen2020imram}
Hui Chen, Guiguang Ding, Xudong Liu, Zijia Lin, Ji~Liu, and Jungong Han. 2020.
\newblock Imram: Iterative matching with recurrent attention memory for cross-modal image-text retrieval.
\newblock In \emph{Proceedings of the IEEE/CVF conference on computer vision and pattern recognition}, pages 12655--12663.

\bibitem[{Chen et~al.(2021)Chen, Hu, Wu, Jiang, and Wang}]{chen2021learning}
Jiacheng Chen, Hexiang Hu, Hao Wu, Yuning Jiang, and Changhu Wang. 2021.
\newblock Learning the best pooling strategy for visual semantic embedding.
\newblock In \emph{Proceedings of the IEEE conference on computer vision and pattern recognition}, pages 15789--15798.

\bibitem[{De~Cao et~al.(2020)De~Cao, Izacard, Riedel, and Petroni}]{de2020autoregressive}
Nicola De~Cao, Gautier Izacard, Sebastian Riedel, and Fabio Petroni. 2020.
\newblock Autoregressive entity retrieval.
\newblock In \emph{International Conference on Learning Representations}.

\bibitem[{Dhariwal and Nichol(2021)}]{dhariwal2021diffusion}
Prafulla Dhariwal and Alexander Nichol. 2021.
\newblock Diffusion models beat gans on image synthesis.
\newblock \emph{Advances in neural information processing systems}, 34:8780--8794.

\bibitem[{Diao et~al.(2021)Diao, Zhang, Ma, and Lu}]{diao2021similarity}
Haiwen Diao, Ying Zhang, Lin Ma, and Huchuan Lu. 2021.
\newblock Similarity reasoning and filtration for image-text matching.
\newblock In \emph{Proceedings of the AAAI conference on artificial intelligence}, volume~35, pages 1218--1226.

\bibitem[{Faghri et~al.(2017)Faghri, Fleet, Kiros, and Fidler}]{faghri2017vse++}
Fartash Faghri, David~J Fleet, Jamie~Ryan Kiros, and Sanja Fidler. 2017.
\newblock Vse++: Improving visual-semantic embeddings with hard negatives.
\newblock \emph{arXiv preprint arXiv:1707.05612}.

\bibitem[{Goodfellow et~al.(2020)Goodfellow, Pouget-Abadie, Mirza, Xu, Warde-Farley, Ozair, Courville, and Bengio}]{goodfellow2020generative}
Ian Goodfellow, Jean Pouget-Abadie, Mehdi Mirza, Bing Xu, David Warde-Farley, Sherjil Ozair, Aaron Courville, and Yoshua Bengio. 2020.
\newblock Generative adversarial networks.
\newblock \emph{Communications of the ACM}, 63(11):139--144.

\bibitem[{Ho et~al.(2020)Ho, Jain, and Abbeel}]{ho2020denoising}
Jonathan Ho, Ajay Jain, and Pieter Abbeel. 2020.
\newblock Denoising diffusion probabilistic models.
\newblock \emph{Advances in neural information processing systems}, 33:6840--6851.

\bibitem[{Huang et~al.(2023)Huang, Dong, Wang, Hao, Singhal, Ma, Lv, Cui, Mohammed, Liu et~al.}]{huang2023language}
Shaohan Huang, Li~Dong, Wenhui Wang, Yaru Hao, Saksham Singhal, Shuming Ma, Tengchao Lv, Lei Cui, Owais~Khan Mohammed, Qiang Liu, et~al. 2023.
\newblock Language is not all you need: Aligning perception with language models.
\newblock \emph{arXiv preprint arXiv:2302.14045}.

\bibitem[{Lee et~al.(2018)Lee, Chen, Hua, Hu, and He}]{lee2018stacked}
Kuang-Huei Lee, Xi~Chen, Gang Hua, Houdong Hu, and Xiaodong He. 2018.
\newblock Stacked cross attention for image-text matching.
\newblock In \emph{Proceedings of the European Conference on Computer Vision}, pages 201--216.

\bibitem[{Lewis et~al.(2020)Lewis, Perez, Piktus, Petroni, Karpukhin, Goyal, K{\"u}ttler, Lewis, Yih, Rockt{\"a}schel et~al.}]{lewis2020retrieval}
Patrick Lewis, Ethan Perez, Aleksandra Piktus, Fabio Petroni, Vladimir Karpukhin, Naman Goyal, Heinrich K{\"u}ttler, Mike Lewis, Wen-tau Yih, Tim Rockt{\"a}schel, et~al. 2020.
\newblock Retrieval-augmented generation for knowledge-intensive nlp tasks.
\newblock \emph{Advances in Neural Information Processing Systems}, 33:9459--9474.

\bibitem[{Li et~al.(2023{\natexlab{a}})Li, Li, Savarese, and Hoi}]{li2023blip}
Junnan Li, Dongxu Li, Silvio Savarese, and Steven Hoi. 2023{\natexlab{a}}.
\newblock Blip-2: Bootstrapping language-image pre-training with frozen image encoders and large language models.
\newblock \emph{arXiv preprint arXiv:2301.12597}.

\bibitem[{Li et~al.(2019)Li, Zhang, Li, Li, and Fu}]{li2019visual}
Kunpeng Li, Yulun Zhang, Kai Li, Yuanyuan Li, and Yun Fu. 2019.
\newblock Visual semantic reasoning for image-text matching.
\newblock In \emph{Proceedings of the IEEE Conference on Computer Vision and Pattern Recognition}, pages 4654--4662.

\bibitem[{Li et~al.(2023{\natexlab{b}})Li, Yang, Wang, Wei, and Li}]{li2023generative}
Yongqi Li, Nan Yang, Liang Wang, Furu Wei, and Wenjie Li. 2023{\natexlab{b}}.
\newblock Generative retrieval for conversational question answering.
\newblock \emph{Information Processing \& Management}, 60(5):103475.

\bibitem[{Li et~al.(2023{\natexlab{c}})Li, Yang, Wang, Wei, and Li}]{li2023learning}
Yongqi Li, Nan Yang, Liang Wang, Furu Wei, and Wenjie Li. 2023{\natexlab{c}}.
\newblock Learning to rank in generative retrieval.
\newblock \emph{arXiv preprint arXiv:2306.15222}.

\bibitem[{Li et~al.(2023{\natexlab{d}})Li, Yang, Wang, Wei, and Li}]{li2023multiview}
Yongqi Li, Nan Yang, Liang Wang, Furu Wei, and Wenjie Li. 2023{\natexlab{d}}.
\newblock Multiview identifiers enhanced generative retrieval.
\newblock In \emph{Proceedings of the 61st Annual Meeting of the Association for Computational Linguistics (Volume 1: Long Papers)}, pages 6636--6648. ACL.

\bibitem[{Lin et~al.(2014)Lin, Maire, Belongie, Hays, Perona, Ramanan, Doll{\'a}r, and Zitnick}]{lin2014microsoft}
Tsung-Yi Lin, Michael Maire, Serge Belongie, James Hays, Pietro Perona, Deva Ramanan, Piotr Doll{\'a}r, and C~Lawrence Zitnick. 2014.
\newblock Microsoft coco: Common objects in context.
\newblock In \emph{Computer Vision--ECCV 2014: 13th European Conference, Zurich, Switzerland, September 6-12, 2014, Proceedings, Part V 13}, pages 740--755. Springer.

\bibitem[{Liu et~al.(2023)Liu, Li, Wu, and Lee}]{liu2023visual}
Haotian Liu, Chunyuan Li, Qingyang Wu, and Yong~Jae Lee. 2023.
\newblock Visual instruction tuning.
\newblock \emph{arXiv preprint arXiv:2304.08485}.

\bibitem[{Qu et~al.(2020)Qu, Liu, Cao, Nie, and Tian}]{qu2020context}
Leigang Qu, Meng Liu, Da~Cao, Liqiang Nie, and Qi~Tian. 2020.
\newblock Context-aware multi-view summarization network for image-text matching.
\newblock In \emph{Proceedings of the 28th ACM International Conference on Multimedia}, pages 1047--1055.

\bibitem[{Qu et~al.(2021)Qu, Liu, Wu, Gao, and Nie}]{qu2021dynamic}
Leigang Qu, Meng Liu, Jianlong Wu, Zan Gao, and Liqiang Nie. 2021.
\newblock Dynamic modality interaction modeling for image-text retrieval.
\newblock In \emph{Proceedings of the 44th International ACM SIGIR Conference on Research and Development in Information Retrieval}, pages 1104--1113.

\bibitem[{Radford et~al.(2021)Radford, Kim, Hallacy, Ramesh, Goh, Agarwal, Sastry, Askell, Mishkin, Clark et~al.}]{radford2021learning}
Alec Radford, Jong~Wook Kim, Chris Hallacy, Aditya Ramesh, Gabriel Goh, Sandhini Agarwal, Girish Sastry, Amanda Askell, Pamela Mishkin, Jack Clark, et~al. 2021.
\newblock Learning transferable visual models from natural language supervision.
\newblock In \emph{International conference on machine learning}, pages 8748--8763. PMLR.

\bibitem[{Radford et~al.(2019)Radford, Wu, Child, Luan, Amodei, Sutskever et~al.}]{radford2019language}
Alec Radford, Jeffrey Wu, Rewon Child, David Luan, Dario Amodei, Ilya Sutskever, et~al. 2019.
\newblock Language models are unsupervised multitask learners.
\newblock \emph{OpenAI blog}, 1(8):9.

\bibitem[{Rasley et~al.(2020)Rasley, Rajbhandari, Ruwase, and He}]{rasley2020deepspeed}
Jeff Rasley, Samyam Rajbhandari, Olatunji Ruwase, and Yuxiong He. 2020.
\newblock Deepspeed: System optimizations enable training deep learning models with over 100 billion parameters.
\newblock In \emph{Proceedings of the International Conference on Knowledge Discovery \& Data Mining}, pages 3505--3506.

\bibitem[{Sutskever et~al.(2014)Sutskever, Vinyals, and Le}]{sutskever2014sequence}
Ilya Sutskever, Oriol Vinyals, and Quoc~V Le. 2014.
\newblock Sequence to sequence learning with neural networks.
\newblock \emph{Advances in neural information processing systems}, 27.

\bibitem[{Tay et~al.(2022)Tay, Tran, Dehghani, Ni, Bahri, Mehta, Qin, Hui, Zhao, Gupta et~al.}]{tay2022transformer}
Yi~Tay, Vinh~Q Tran, Mostafa Dehghani, Jianmo Ni, Dara Bahri, Harsh Mehta, Zhen Qin, Kai Hui, Zhe Zhao, Jai Gupta, et~al. 2022.
\newblock Transformer memory as a differentiable search index.
\newblock \emph{arXiv preprint arXiv:2202.06991}.

\bibitem[{Touvron et~al.(2023{\natexlab{a}})Touvron, Lavril, Izacard, Martinet, Lachaux, Lacroix, Rozi{\`e}re, Goyal, Hambro, Azhar et~al.}]{touvron2023llama}
Hugo Touvron, Thibaut Lavril, Gautier Izacard, Xavier Martinet, Marie-Anne Lachaux, Timoth{\'e}e Lacroix, Baptiste Rozi{\`e}re, Naman Goyal, Eric Hambro, Faisal Azhar, et~al. 2023{\natexlab{a}}.
\newblock Llama: Open and efficient foundation language models.
\newblock \emph{arXiv preprint arXiv:2302.13971}.

\bibitem[{Touvron et~al.(2023{\natexlab{b}})Touvron, Martin, Stone, Albert, Almahairi, Babaei, Bashlykov, Batra, Bhargava, Bhosale et~al.}]{touvron2023llama2}
Hugo Touvron, Louis Martin, Kevin Stone, Peter Albert, Amjad Almahairi, Yasmine Babaei, Nikolay Bashlykov, Soumya Batra, Prajjwal Bhargava, Shruti Bhosale, et~al. 2023{\natexlab{b}}.
\newblock Llama 2: Open foundation and fine-tuned chat models.
\newblock \emph{arXiv preprint arXiv:2307.09288}.

\bibitem[{Van Den~Oord et~al.(2017)Van Den~Oord, Vinyals et~al.}]{van2017neural}
Aaron Van Den~Oord, Oriol Vinyals, et~al. 2017.
\newblock Neural discrete representation learning.
\newblock \emph{Advances in neural information processing systems}, 30.

\bibitem[{Young et~al.(2014)Young, Lai, Hodosh, and Hockenmaier}]{young2014image}
Peter Young, Alice Lai, Micah Hodosh, and Julia Hockenmaier. 2014.
\newblock From image descriptions to visual denotations: New similarity metrics for semantic inference over event descriptions.
\newblock \emph{Transactions of the Association for Computational Linguistics}, 2:67--78.

\bibitem[{Zhang et~al.(2023)Zhang, Zhang, Chen, Wang, Chen, Xie, Sun, Deng, Zhang, Yang et~al.}]{zhang2023irgen}
Yidan Zhang, Ting Zhang, Dong Chen, Yujing Wang, Qi~Chen, Xing Xie, Hao Sun, Weiwei Deng, Qi~Zhang, Fan Yang, et~al. 2023.
\newblock Irgen: Generative modeling for image retrieval.
\newblock \emph{arXiv preprint arXiv:2303.10126}.

\bibitem[{Zheng et~al.(2020)Zheng, Zheng, Garrett, Yang, Xu, and Shen}]{zheng2020dual}
Zhedong Zheng, Liang Zheng, Michael Garrett, Yi~Yang, Mingliang Xu, and Yi-Dong Shen. 2020.
\newblock Dual-path convolutional image-text embeddings with instance loss.
\newblock \emph{ACM Transactions on Multimedia Computing, Communications, and Applications (TOMM)}, 16(2):1--23.

\bibitem[{Zhu et~al.(2023)Zhu, Chen, Shen, Li, and Elhoseiny}]{zhu2023minigpt}
Deyao Zhu, Jun Chen, Xiaoqian Shen, Xiang Li, and Mohamed Elhoseiny. 2023.
\newblock Minigpt-4: Enhancing vision-language understanding with advanced large language models.
\newblock \emph{arXiv preprint arXiv:2304.10592}.

\end{thebibliography}
\appendix
\section{Implement Details}
\label{sec: Implement Details}
We selected the open-flamingo~\cite{awadalla2023openflamingo} with the 3B parameters as our model's backbone. The visual encoder of the open-flamingo is a 12-layer visual transformer, while its language model is based on MPT-1B\footnote{\url{https://huggingface.co/anas-awadalla/mpt-1b-redpajama-200b}.}. We adopted the deepspeed~\cite{rasley2020deepspeed} training framework to train the model on 4$\times$24GB NVIDIA A5000 GPUs. We froze the visual encoder and fine-tuned the language model as well as cross-attention layers. We employed the Adam optimizer, setting a learning rate of 1e-4 and a batch size of 64 for each GPU. On the Flickr30K dataset, our training included 1,000K steps for learning to memorize and 3,000K steps for learning to retrieve. For the MS-COCO dataset, these numbers were increased to 2,000K and 6,000K steps, respectively. We have trained the model several times to confirm that the improvement is not a result of random chance and present the mid one. The training duration was approximately 12 hours for Flickr30K and 24 hours for MS-COCO.
\section{Beam size analysis}
\label{sec: Beam size analysis}
\begin{figure}[h]
\centering
\includegraphics[width=0.47\linewidth]{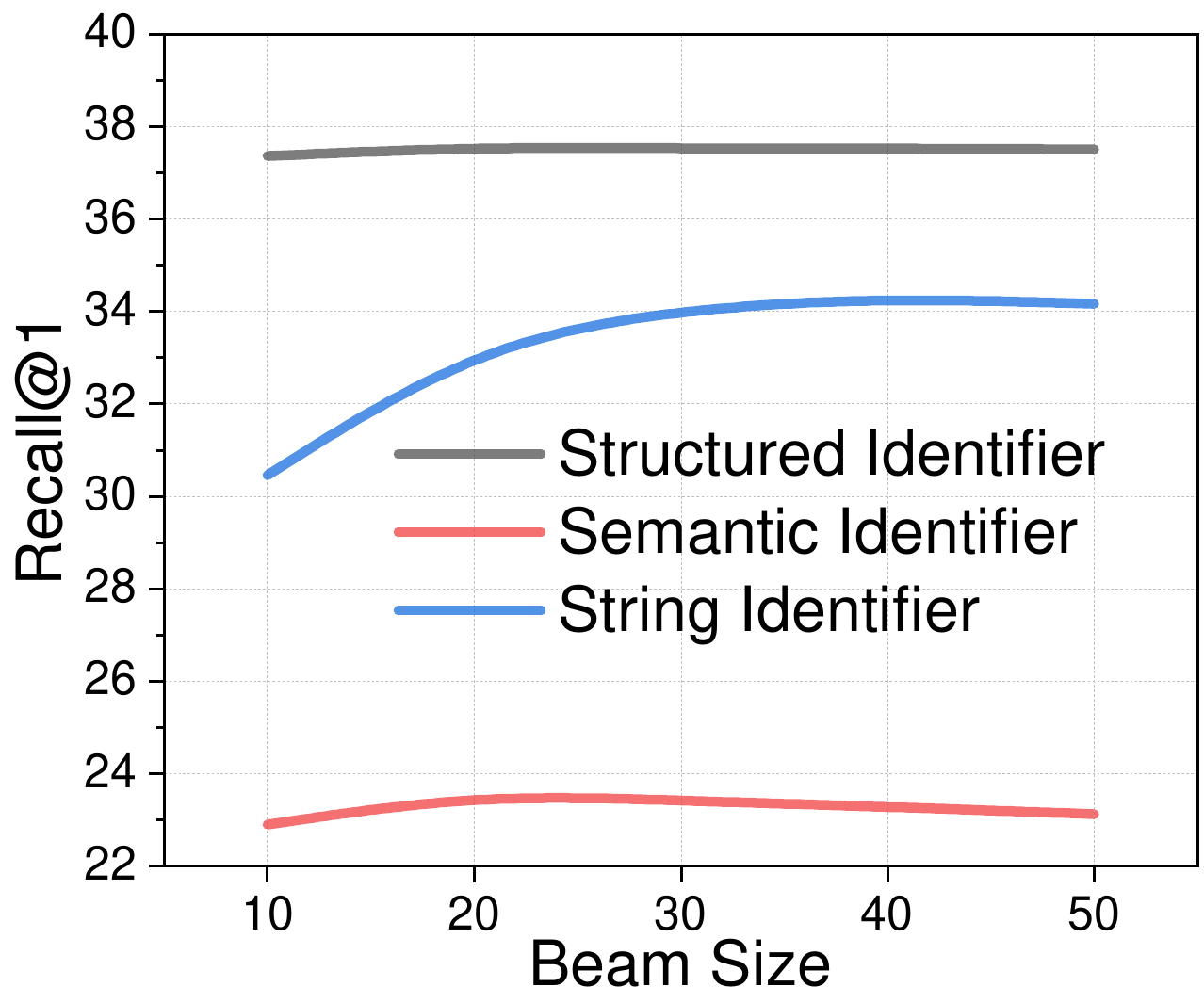}
\label{beam_size:subfig1}
\includegraphics[width=0.47\linewidth]{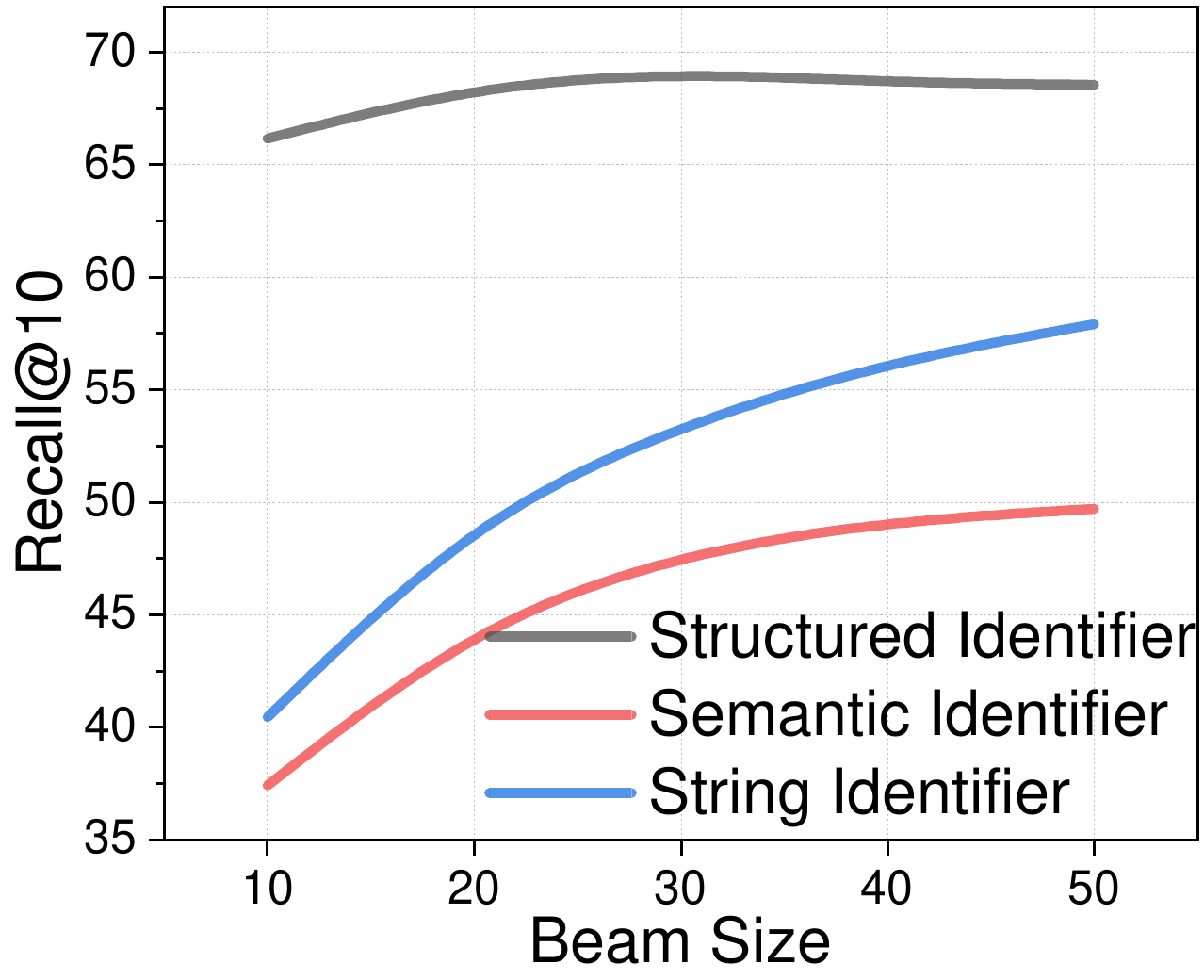}
\label{beam_size:subfig2}
\caption{The retrieval performance of different identifiers varies by beam sizes in terms of Recall@1 and Recall@10.}
\label{beam_size}
\end{figure}

GRACE relies on beam search to obtain top-k retrieval results. We conducted detailed experiments to understand the impact of varying beam sizes, and the findings are illustrated in Figure~\ref{beam_size}. The atomic identifier is excluded from this experiment as it only requires one generation step, and beam size will not affect its performance.

Increasing the beam size exhibits marginal benefits. This observation aligns with our expectations, as candidates with larger beam sizes generally score lower, diminishing their likelihood of being the top result. In terms of Recall@10, we observed a notable improvement in performance with the expansion of the beam size. This enhancement is attributed to the inclusion of candidates that would otherwise be missed in scenarios with a more constrained beam size.

\end{document}